\documentclass[conference]{IEEEtran}
\IEEEoverridecommandlockouts
\usepackage{cite}
\usepackage{amsmath,amssymb,amsfonts}
\usepackage{algorithmic}
\usepackage{graphicx}
\usepackage{textcomp}
\usepackage{mathtools}
\usepackage{booktabs}

\usepackage{color}
\usepackage{xcolor}

\def\BibTeX{{\rm B\kern-.05em{\sc i\kern-.025em b}\kern-.08em
    T\kern-.1667em\lower.7ex\hbox{E}\kern-.125emX}}

\begin{document}

\title{PDF-Malware: An Overview on Threats, Detection and Evasion Attacks}

\author{
\IEEEauthorblockN{Nicolas Fleury}
\IEEEauthorblockA{
\textit{Université Polytechnique }\\
\textit{Hauts-De-France} \\Valenciennes, France \\
nicolas.fleury@etu.uphf.fr
}
\and

\IEEEauthorblockN{Theo Dubrunquez}
\IEEEauthorblockA{
\textit{Université Polytechnique }\\
\textit{Hauts-De-France} \\Valenciennes, France \\
theovalentin.dubrunquez@etu.uphf.fr}

\and
\IEEEauthorblockN{Ihsen Alouani}
\IEEEauthorblockA{\textit{IEMN-DOAE Lab CNRS 8520} \\
\textit{Université Polytechnique Hauts-De-France}\\
Valenciennes, France \\
ihsen.alouani@uphf.fr}


}

\maketitle

\begin{abstract}
In the recent years, Portable Document Format, commonly known as PDF, has become a democratized standard for document exchange and dissemination. This trend has been due to its characteristics such as its flexibility and portability across platforms. The widespread use of PDF has installed a false impression of inherent safety among benign users. However, the characteristics of PDF motivated hackers to exploit various types of vulnerabilities, overcome security safeguards, thereby making the PDF format one of the most efficient malicious code attack vectors. Therefore, efficiently detecting malicious PDF files is crucial for information security. Several analysis techniques has been proposed in the literature, be it static or dynamic, to extract the main features that allow the discrimination of malware files from benign ones. Since classical analysis techniques may be limited in case of zero-days, machine-learning based techniques have emerged recently as an automatic PDF-malware detection method that is able to generalize from a set of training samples. These techniques are themselves facing the challenge of evasion attacks where a malicious PDF is transformed to look benign. In this work, we give an overview on the PDF-malware detection problem. We give a perspective on the new challenges and emerging solutions.
\end{abstract}


\section{Introduction }

   

Malicious attackers compromise systems to install malware \cite{rhmd33, rhmd48} to gain access and privilege, to compromise personal or sensitive data, to sabotage systems, or to use them in other attacks such as DDOS \cite{ddos}. Preventing the compromise of information systems is practically impossible. In fact, attackers succed the intrusions in a variety of manners, such as drive-by-downloads with websites exploiting browser vulnerabilities \cite{rhmd2} or network-accessible vulnerabilities \cite{rhmd46}. Besides, social engineering attacks such as Phishing attacks, and malicious email attachments allow user-authorized installation of malicious binaries \cite{rhmd1}.

\begin{figure}
\begin{center}
    \includegraphics[width=\columnwidth]{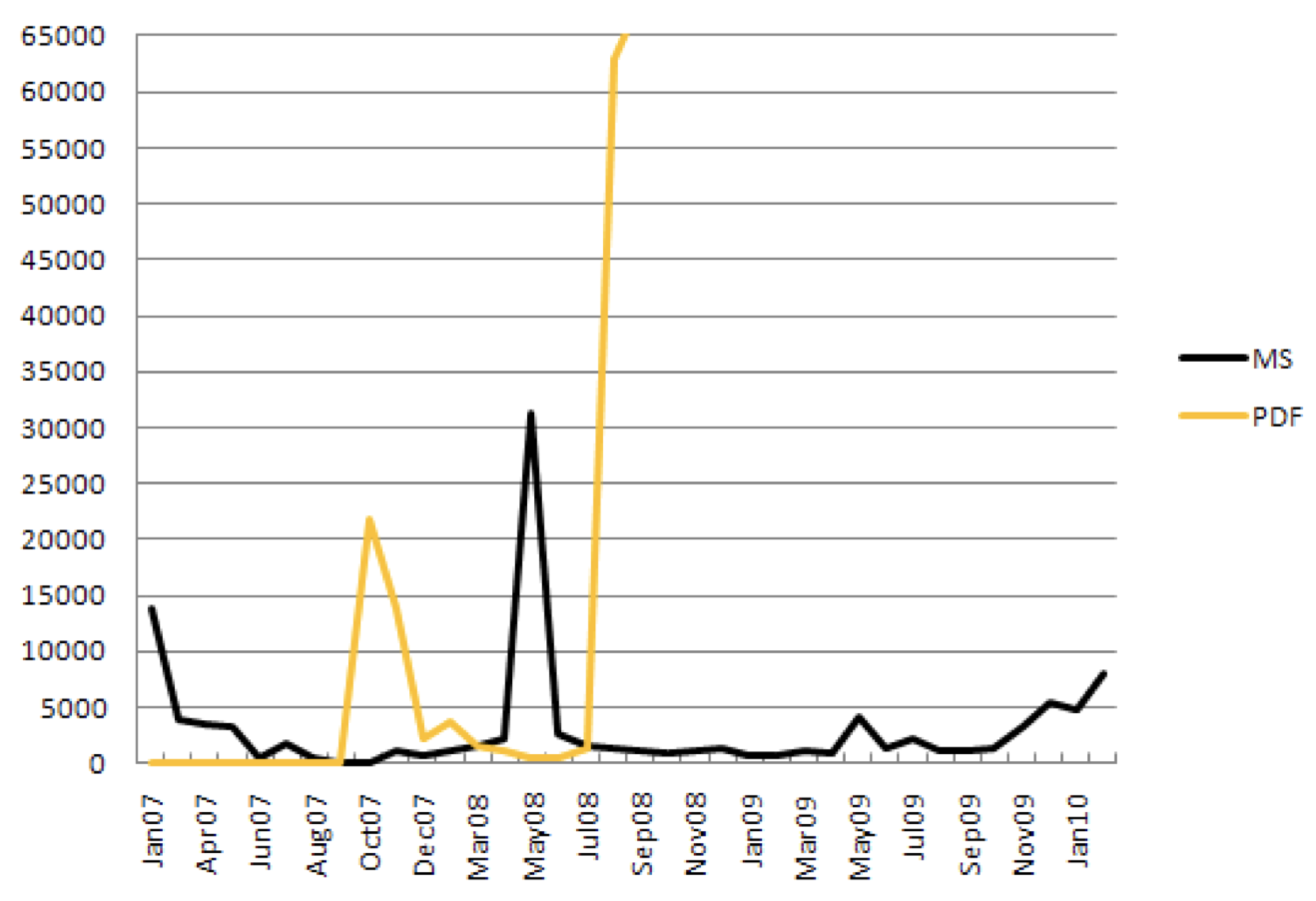}
    \caption{Number of attacks: Microsoft Office vs. PDF \cite{symantec}.} 
\end{center}

  \label{fig:symantec}
\end{figure}

Regular end users are easily able to see the threat of a clear binary and executable files. Their awareness is also increasing against many threat vectors such as Microsoft Office documents including macros. However, despite the complexity PDF format, end users still tend to consider that PDF files are harmless static documents. This implicit assumption mainly results from ignoring the fact that what PDF file displays is the execution output of a potentially complex program; mainly javascript code running in the background. In 2010 Symantec \cite{symantec}  reported a large rise in PDF-driven attacks, mainly justifying it with a corresponding rise in the vulnerabilities identified in the Adobe Reader software. More recently, Ke Liu reported \cite{Keliu} about his discovery since December 2015 of more than 150 vulnerabilities in the most common PDF reader software products. This latter news shows how, even today, PDF is an important infection vector that provides a large attack surface. As shown in Figure \ref{fig:symantec}, the amount of PDF attacks Symantec has
recorded \cite{symantec} have increased dramatically, which shows that the PDF file format is being targeted more often. The spikes on these graphs coincide with the release of specific PDF-related CVEs. 

The potential attack vector of PDF files combined with a widespread wrong assumption of harmlessness makes the detection of malicious PDF an important topic for the information security community.

Malware developers typically exploit the possibility to supply Javascript to the PDF reader interpretation engine to execute malicious code. Such code is usually sandboxed for execution, but it may still exploit unpatched vulnerabilities to escape the environment boundaries and execute shellcode at the user level. Complex payloads can be included in the PDF as obfuscated text to evade inspection techniques, or can be downloaded from the Internet as soon as the attacker takes control of the user shell. Malicious PDF files are then delivered through different methods \cite{symantec}: from drive-by downloads, to targeted attacks or mass mailing approaches.

This paper aims at presenting a brief overview on the main PDF-malware threats, the main detection techniques and gives a perspective on emerging challenges in detecting PDF-malware. 

The remainder of the paper is organized as follows: Section 2 presents a brief background on PDF format as well as on machine learning. Section 3 presents the PDF-based threat used by attackers. Section 4 gives an overview on state of the art malware detection techniques. The evasion attacks challenge is explained in Section 5. We finally give concluding remarks in Section 6.

\section{ Background}
\subsection{The Portable Document Format}

\begin{figure*}
  
\begin{center}
    \includegraphics[width=1.3\columnwidth]{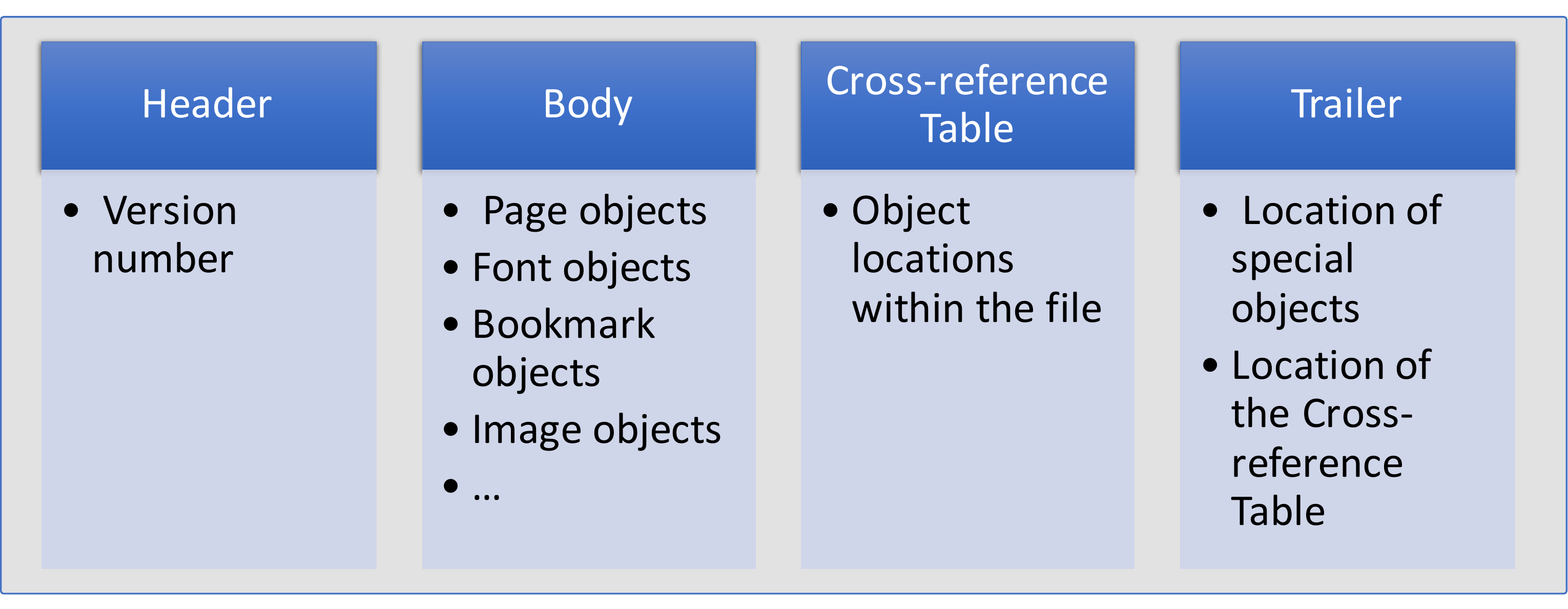}
    \caption{Simplified structure of a PDF file.} 
    \label{fig:pdf1}
\end{center}
\end{figure*}

The Portable Document Format is the world’s most widely used for both paper and online format for printed documents. PDF was defined in 1993 by Adobe Systems and used until today to exchange and print documents regardless the underlying hardware architecture, software platform, and operating system. In 2008, PDF became an open standard released as ISO 32000-1.
A PDF file may contain a mix of textual and binary data and is composed by different abstraction layers. The layers define the sequential flow by which a PDF viewer application reads the contents and renders them on the screen. According to the PDF Reference \cite{pdf}, the internal structure of a PDF file is made up of the elements depicted in Figure \ref{fig:pdf1}.

The PDF contains four sections: header, body, cross-reference table and trailer. 

\begin{itemize}
    \item The header is used to identify the file format and the version, $\%PDF-1.x$ where $x$ is a number between 0 and 7. However, the header could be placed anywhere in the first 1024 bytes. If PDF file contains data, that line is followed by a comment line containing at least four binary characters whose codes are 128 or greater.
    \item The body contains multiple types of objects, and these are the most important: 
     \textbf{(i) Objects:} They may be either direct (embedded in another object) or indirect. Indirect objects are identified with an object number and a generation number (object's version number) and defined between the $obj$ and $endobj$ keywords if residing in the document root.\\
      \textbf{(ii) A dictionary:} object that starts with "\textless\textless" and end with "\textgreater\textgreater" and is enclosed by $obj$ and $endobj$ keywords.\\
     \textbf{(iii)  A stream object:} it is represented by a sequence of bytes and may be unlimited in length, which is why images, javascript and other big-size data blocks are usually represented as streams. Stream object can make use of a special feature called filters. Filters can be used for different purposes such as encoding or decoding of content, compression and decompression. Furthermore, multiple filters can be applied on a stream object. 

    \item The cross-reference table: it indexes all objects' locations in the file. This table can have multiple subsection containing objects, represented by 2 numbers : the first number corresponds to the object number, while the second line states the number of objects in the current subsection, so if the object number is $0$ and we have $3$ objects, we will have objects $0, 1$ and $2$. Objects are represented by one entry, which is 20 bytes: 10 first bytes are the object offset from the start of the PDF document to the beginning of that object, followed by a space separator with another number specifying the object's generation number. After that there is an other space separator followed by a letter $f$ or $n$ indicating if the object is free or in use.
    \item The trailer is the first thing to be processed in a PDF and it specifies how the application reading the PDF document should find the cross-reference table and other special objects. The trailer's dictionary generally contains the document's catalogue object, and sometimes the document's information dictionary in which we can find the creation and modification dates of the file, together with some simple metadata.
\end{itemize} 

 An important but critical feature of PDF comes from the fact that a document can be modified or updated in an incremental way. This means that if a file is updated by adding a new body, cross-reference table and trailer without changing anything in the rest of the file. That feature allows any user coming back with original data by cancelling the modifications.



\subsection{Machine Learning}

\begin{figure}[!htp]
\begin{center}
    \includegraphics[width=0.7\columnwidth]{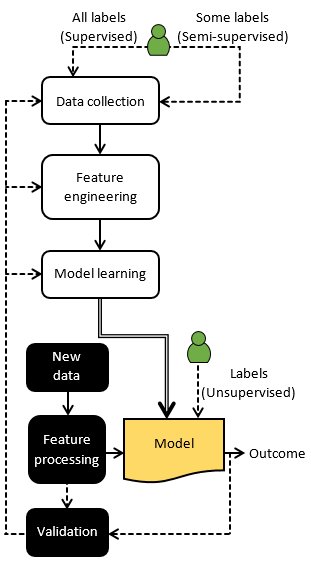}
    \caption{Design methodologies of different machine learning categories \cite{articleML}.} 
    \label{fig:MLCat}
\end{center}
\end{figure}

Machine Learning (ML) is one of the most useful tools nowadays. It has shown in recent years an impressive capability to effectively deal with a plethora of  complex real-life problems. The main characteristic in ML computing paradigm is to create knowledge from data. A ML algorithm goes through a training phase with a dataset until converging to a \emph{trained} state where it can be tested and then validated on a separate data set. A trained model is expected to be able to generalize to unseen samples. As shown in Figure \cite{articleML}, the training depends mostly on the available data structure: it can be supervised when data is labeled, unsupervised when no labels are available, or semi-supervised when data is partially labeled. 

There is four important steps in ML design. The First is to determine the category that suits the problem. There are four main categories: clustering, classification, regression and rule extraction. Secondly, once the category is fixed, a specific model corresponding to the category needs to be identified. For example, one could choose Artificial Neural Network, Random Forest, Naive Bayesian, a Support Vector Machine (SVM), etc. Then, using the available data set, the model goes through a training process to identify the optimal parameters of the model that solves the considered problem. The final step is to test the model with data that has not been seen in training process. This step is also important because it’s where we will get all the metrics to validate or not the model. The accuracy and false negative rate are the more representative of the model but there is also others such as average precision, specificity, F1-score, etc.

\section{PDF Malware}


It is important to know that PDF can be a great attack vector because a lot of people believes it's safe and don't even suspect a PDF to be potentially dangerous. Email attachments combined with social engineering are among many attack vectors cybercriminals take advantage of. In addition to email attachments, the use of web malware exploitation is one of the most widely used attack vectors. 

There are different ways to perform malicious actions using PDF. The most common attack vector for malware PDFs derives from embedded JavaScript code that can be executed by the PDF reader. Indeed, many surveyed papers consider features derived in different ways from embedded JavaScript code \cite{surv_2, surv_4, surv_7, surv_9, surv_19, surv_25}.

The following are some well-known PDF-based attack scenarios:
    \begin{itemize}
    
        \item  OpenAction feature can be used to set an exploit when the file is opened. An action is a legitimate PDF feature. Some potentially dangerous actions include Launch, Go-to, Universal Resource Indicator (URI), Named and JavaScript actions \cite{tech_malw, rec_pattern}.
    
        \item Launch action,  giving the possibility to launch special commands on the operating system, and could run an executable if the user clicks OK on the confirmation windows that is opened \cite{tech_malw}.
        
        \item  Embedded files, which can be extracted and opened by the reader. This may be used to hide malicious executables or malicious PDF, Embedded Flash applications stored as embedded SWF files or malicious ActionScript code \cite{tech_malw, rec_pattern}.
  
        \item  GotoEmbedded action can be dangerous as PDF files can contain embedded PDF files, which can be encrypted. When a user loads the main PDF file, it could immediately load its embedded PDF file. This allows attackers to hide malicious PDF files inside other PDF files, fooling antivirus scanners by preventing them from examining the hidden PDF file \cite{tech_malw, rec_pattern}. 
        
        \item URI action allows access to a remote resources by mean of an Universal Resource Indicator. This way, an attacker could redirect an user to a malicious website \cite{PDFURI} or exfiltrate data \cite{URIexf} by combining that feature with Javascript, OpenAction or using PDF forms (with the Submit Form action).
    \end{itemize}

\section{PDF Malware Detection}

The most commonly used way for detecting PDF malware is to search files for signatures or patterns of known malware. While this widely used techniques in classical anti-virus software is fast and pragmatic, they are easily fooled and overcome by attacker through simple evasion and obfuscation techniques. In fact, in addition to its ineffectiveness against zero-days, even if the vulnerable APIs that malware uses as an attack vector might be known, detecting them syntactically can be evaded by an attacker through obfuscation. Several public datasets are available to develop PDF-malware detection techniques; Contagio \cite{contagio} is one of the most widely used ones.

\begin{figure}
\begin{center}
    \includegraphics[width=0.79\columnwidth]{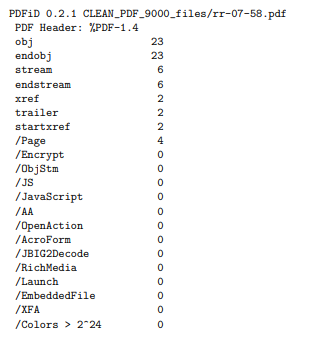}
    \caption{Output of PDFiD \cite{staticML}. }
\end{center}
  \label{fig:PDFID}
\end{figure}

\subsection{Static analysis}
Static analysis can be done by looking directly at the content of the file or using specific tools. PDFiD \cite{stevensBlog} or peepdf \footnote{https://github.com/jesparza/peepdf} are among the most widely used tools to statically analyze PDFs. 
PDFiD is fine if you need a quick overview of what is in your PDF file but if you want a better and deeper analysis peepdf might be a better choice.
PDFiD python script was designed by Didier Stevens \cite{stevensBlog}. This script scans through a PDF file, and counts the number of occurences of each features. These 21 features are commonly found in malicious files. PDFiD gives a simple and fast overview \ref{fig:PDFID} of what the PDF contains (Javascript, Open action, Launch action...)

\subsection{Dynamic analysis}

\begin{figure}
\begin{center}
    \includegraphics[width=0.9\columnwidth]{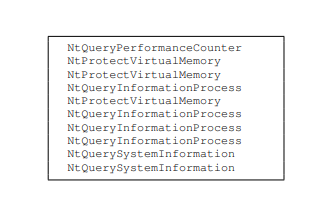}
    \caption{Example system call trace of process (truncated to 10 calls) \cite{7272922}.} 
    \label{fig:syscallEx}
\end{center}

\end{figure}

\begin{figure*}[h]
\begin{center}
    \includegraphics[width=1.4\columnwidth]{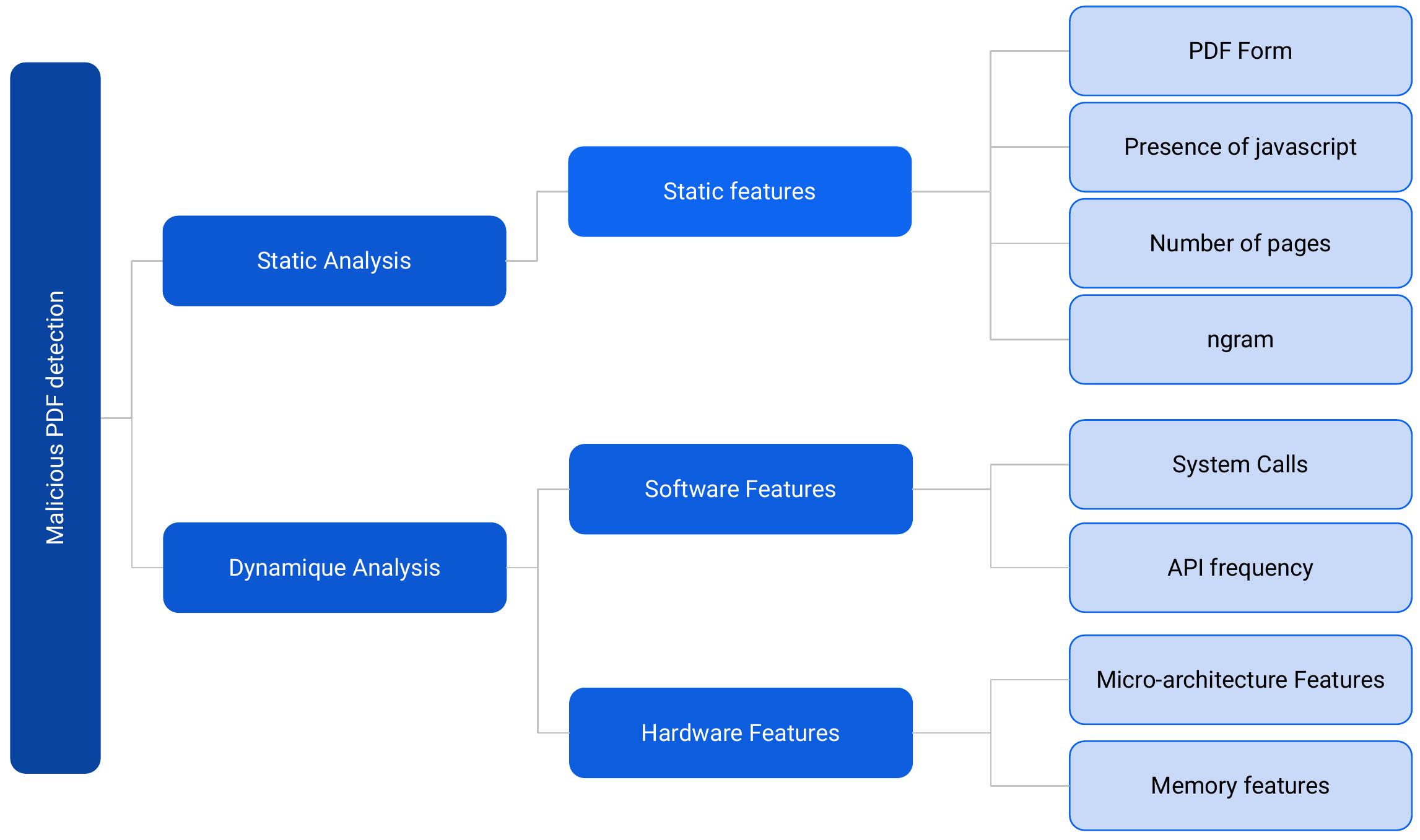}
    \caption{An overview on PDF-malware detection features.} 
     \label{fig:GraphPDF}
\end{center}
\end{figure*}

Unlike the static analysis, dynamic analysis is performed at runtime. One of the challenges that are specific to PDF-malware is the fact that PDF documents are not executable and are launched through a PDF reader. Then, the analysis needs to be performed on a vulnerable machine so that the payload, if any, can be triggered and thereby analysed. If the payload does not run due to security measures, the results are useless. Varying the PDF viewer is also essential since some malicious PDF are made for a specific viewer or even for a specific version of a viewer ( ex: Adobe Acrobat Reader DC ). Once again, the priority with dynamic analysis is to capture what happen when the payload is running. Once this is achieved, the analysis process has to collect APIs and are system calls. These are the main traces that are useful to detect the potentially malicious requests for operating system services \cite{7272922}. An illustration of such traces is given in Figure \ref{fig:syscallEx}.

The tools that you can use are for example strace \footnote{https://github.com/strace} for Linux or dtrace \footnote{https://docs.microsoft.com/en-us/windows-hardware/drivers/devtest/dtrace} under Windows OS. Once traces are collected, postprocessing is needed to make them more human readable; for example by sorting them by type of syscalls. 

\subsection{Hardware Malware Detection}
While most of the existing malware analysis approaches tackle the problem from sofware abstraction level, a number of works have looked at using low-level features. These approaches are referred to as Hardware Malware Detectors (HMDs) and rely on micro-architecture features such as frequency of opcodes \cite{rhmd7}, evaluation of opcode sequence signatures \cite{rhmd45}. These features are collected while a binary is running and analysed for malware behavior detection. In \cite{rhmd44}, offline analysis is performed through opcode sequence similarity graphs.
In the same direction, Demme et al. \cite{rhmd12} proposed collecting performance counter statistics for programs and malware under execution and
used them to show that offline detection of malware is effective.
Then, a real-time hardware malware detector was built by Ozsoy et al. \cite{rhmd39}. Tang et al. \cite{rhmd50} used unsupervised learning to detect
malware exploits, which will make the regular program deviate from the baseline execution model. Kazdagli et al \cite{rhmd25} identified some pitfalls in training and evaluating HMDs for mobile malware, and
proposed several improvements to them.


\subsection{Machine-learning based techniques}
The main goal of using machine learning for malware detection is to build a classifier that is able to detect malicious PDFs that he has never seen. Ideally, that should help to prevents new attacks and that should be more robust than a classic antivirus. One can extract features using static analysis and perform an analysis using an artificial neural network.
    
As Explained in section IV.A. PDFiD can be used to extract features to train a model. For example related work \cite{staticML} implemented this solution. They used a dataset of $10 000$ clean and $10 000$ malicious PDF documents. The model was a SVM implemented in Python using $60\%$ of the dataset for training and the other 40\% for test. They obtain an accuracy of $99.60\%$ and a false-positive rate of $0.05\%$.

Notice that static, dynamic, software and hardware features can be used to design a ML-based PDF-malware detection system. 

\section{Evasion Attacks}
\subsection{Attack Mechanism}

The main goal of these Evasion attacks is to fool the classifier by changing the features of the infected PDF files so that the classifier considers them as clean. To have an effective attack, these modifications should not be noticeable by the defender by scanning the appearance of the file. Removing objects is not effective for evasion because most of the time, it will change the behavior and probably the display of the file. On the other hand, adding empty objects seems to be a good and easy way to modify a PDF file without damaging its original content.
We consider a white box adversary. In this model, the adversary has access to everything the defender has, namely: the training dataset used to train the classifier, the classifier algorithm, the classifier parameters (kernel, used features for vector, etc.), and infected PDF files that are detected by the classifier. 
We experimented an Evasive attack against a Classifier (ANN) we trained on Contagio dataset \cite{contagio}. Our intuition was that, in our dataset and in general, infected PDF only contain the payload with no more content. Hence, a simple solution was to add enough object in an infected PDF to make it look like  a normal PDF (in term of number objects).
Most of the PDF readers were able to find PDF file's objects even if the objects location in the cross-reference are wrong. This means that adding objects in a PDF file is easy and doesn't affect the PDF behavior in most of the cases. We implemented this attack to generate evasive examples and we obtained more than $98\%$ attack success rate, i.e., the classifier was not able to recognize the evasive malware.
A very similar attack has already been implemented \cite{staticML}, the attacker picks one feature and increments it until the vector is considered as clean by the classifier.

Other techniques inspired from adversarial attacks in image applications are based on gradient-descent to analytically find the minimum noise needed to fool the system. It has been used to evade Support Vector Machines (SVMs) and neural networks classifiers \cite{staticML,gdAtt}. Moreover, this approach is applicable to any classifier with a differentiable discriminant function.

\subsection{Defenses \& Perspectives}
A counter-measure that can be applied to counter the first attack we proposed is to use a maximum value for our features, and that totally blocks the attack when that value was set up to 1.
The gradient-descent attack works very well because the algorithm has a huge degree of freedom due to the possibility of increasing every component of the vector as much as required. Selecting robust features could be a solution, but that would a deeper analysis of the PDF \cite{robustFeatures} 
and use a threshold for our features could counter the gradient-descent. \cite{staticML}

However, we could simply train our classifier using the files we used to evade it : it is called adversarial learning \cite{adv:train}. 

In previous work \cite{fcombim}, combination of static and dynamic features seems to improve the detection rate of malicious Mobile App, and we think that it is worth exploring to utilize it in PDF-malware context. We believe that combining static, dynamic and hardware features can enhance the classifier robustness against evasion attacks.

\section{Conclusion}
In this paper, we present a brief overview on the main PDF-malware threats, the main detection techniques. We give a perspective on emerging challenges in detecting PDF-malware and suggest ideas to enhance PDF malware detectors robustness.


\bibliographystyle{IEEEtran}
\bibliography{biblio}


\end{document}